\documentclass[12pt]{article}                                                   
\usepackage{graphicx}

\newcommand{\gsim}{\hbox{ \raise3pt\hbox to 0pt{$>$}\raise-3pt\hbox{$\sim$} }}
\newcommand{\lsim}{\hbox{ \raise3pt\hbox to 0pt{$<$}\raise-3pt\hbox{$\sim$} }}
\newcommand{\del}{\ifmmode{\nabla}               \else{$\nabla$ }               \fi}

\begin{document}

\begin{center}
{\Large \bf A Beam Test of Prototype TPCs \\
using Micro-Pattern Gas Detectors at KEK \\}
\vspace{3mm}
\small{\bf
-----
An Interpretation of the Results and Extrapolation to the ILC-TPC
-----} \\
\vspace{5mm}
Makoto Kobayashi \\
----- On behalf of part of the ILC-TPC Collaboration ----- \\
\vspace{3mm}
{\footnotesize High Energy Accelerator Organization (KEK), \\ 
Tsukuba 305-0801, Japan}
\end{center}
\begin{abstract}
We conducted a series of beam tests of prototype TPCs for the International
Linear Collider (ILC) experiment, equipped with an MWPC, a MicroMEGAS, or GEMs
as a readout device.
The prototype operated successfully in a test beam at KEK under an axial
magnetic field of up to 1 T.
The analysis of data is now in progress and some of the preliminary results
obtained with GEMs and MicroMEGAS are presented along with our interpretation.
Also given is the extrapolation of the obtained spatial resolution
to that of a large TPC expected for the central tracker of the
ILC experiment.
\end{abstract}
\vspace{1mm}
\hspace{8.5mm} {\bf keywords.}  {\small TPC; MPGD; MicroMEGAS; GEM; ILC;
                               Spatial Resolution.}
\vspace{1mm}
\section{Introduction}
One of the major physics goals of the future linear collider experiment
is to study properties of the Higgs boson, which is expected to be well
within the reach of the center-of-mass energy of the machine
~\cite{ILC1}~\cite{ILC2}.
This goal demands unprecedented high performance of each detector component.
For example, the central tracker is required to have a high momentum
resolution, high two-track resolving power, and a high momentum resolution,
for precise reconstruction of hard muons and each of charged particle tracks
in dense jets.

A time projection chamber (TPC) is a strong candidate for the central tracker
of the experiment since it can cover a large volume with a small material
budget while maintaining a high tracking density (granularity).
If micro-pattern gas detectors (MPGDs: micro-mesh gaseous structure
(MicroMEGAS)~\cite{Giomataris}, gas electron multiplier (GEM)~\cite{Sauli} etc.)
are employed for the detection devices of the TPC, instead of conventional
multi-wire proportional chambers (MWPCs),
one can expect a better spatial resolution at a lower gas gain, a higher
granularity, and a smaller or negligible $E \times B$ effect at the entrance
to the detection plane.
Furthermore, the MPGDs have inherently smaller positive-ion back flow rate
than that of MWPCs.
We therefore constructed a small prototype TPC with a replaceable readout
device (MWPC, MicroMEGAS or triple GEM) and have conducted a series of beam
tests at KEK in order to study the performance, especially its spatial
resolution under an axial magnetic field.

We begin with brief descriptions of the prototype TPC and the experimental
setup.
Next, some preliminary results are presented along with our interpretation,
in which special emphasis is placed on an analytic expression of the spatial
resolution.
Finally, the spatial resolution of the ILC-TPC is estimated from that 
measured with the prototype. \\   
------------------------------------------------------\\
{\footnotesize \boldmath $Contributed~paper~to~the~Linear~Collider~Workshop,
~March~2006,~I.I.Sc~Bangalore,~India$} \\

\section{Experimental setup}

A photograph of the prototype is shown in Fig.~1.
\begin{figure}[htbp]
\begin{center}
\hspace{10mm}
\includegraphics[width=14.0cm,clip]{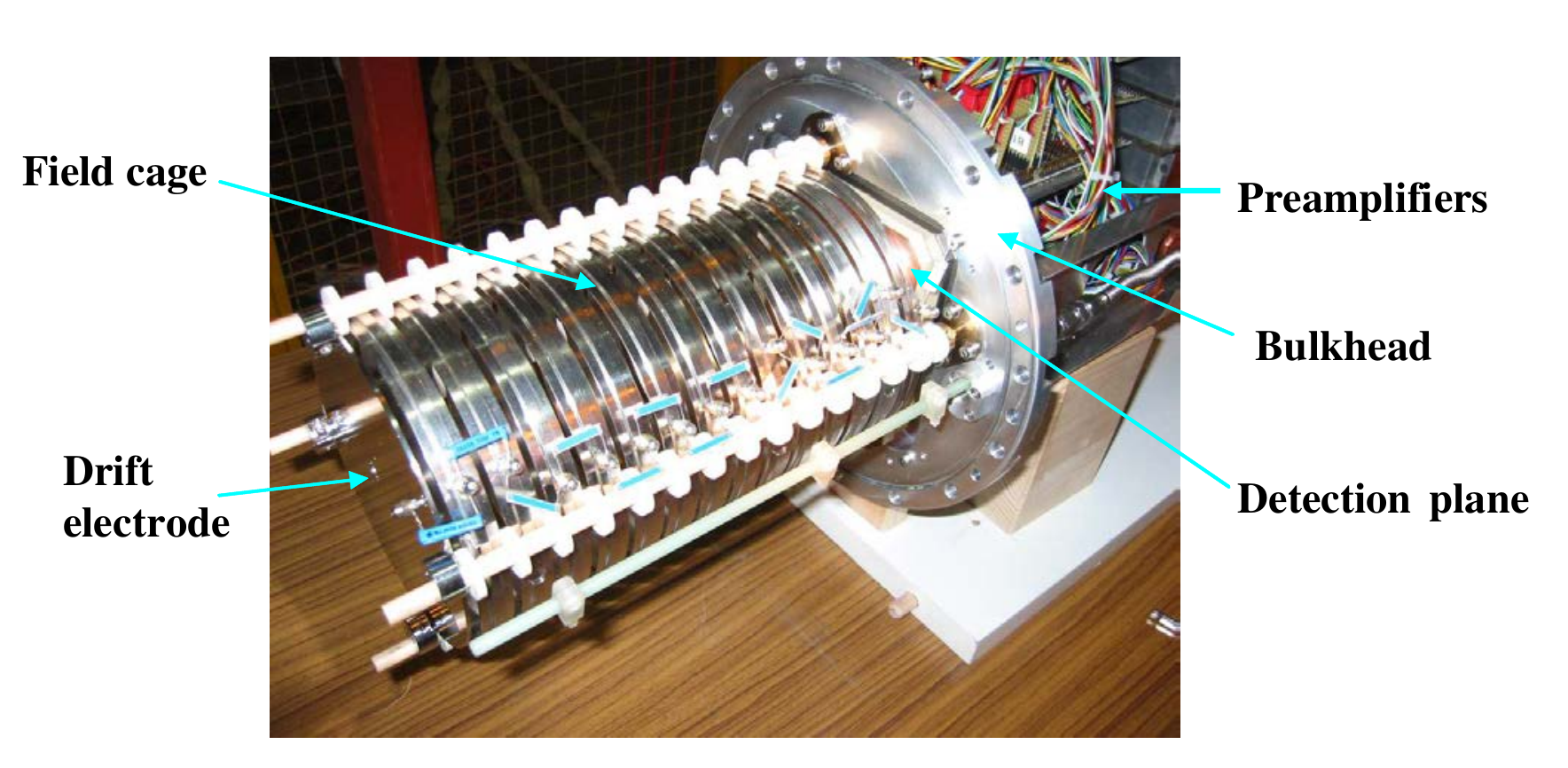}
\caption[fig1]{\label{fig1}
\footnotesize Photograph of the prototype just before installation
into the gas vessel.
}
\end{center}
\end{figure}
It consists of a field cage and an easily replaceable gas amplification
device attached to one end of the field cage.
Gas amplified electrons are detected by a pad plane at ground potential
placed right behind the amplification device.
A drift electrode is attached to the other end of the field cage.
The maximum drift length is about 260 mm.

The pad plane, with an effective area of $\sim 75 \times 75$ mm$^2$,
has 12 pad rows at a pitch of 6.3 mm, each consisting of
$2 \times 6$ ($1.27 \times 6$) mm$^2$ rectangular pads arranged at a pitch of
2.3 (1.27) mm when combined with MicroMEGAS (GEMs).
Pad signals are fed to charge sensitive preamplifiers located on the outer
surface of the bulkhead of the gas vessel behind the pad plane.
The amplified signals are sent to shaper amplifiers with a shaping time of
500 ns in the counting room via coaxial cables, and then processed by 
12.5 MHz digitizers.

The mesh of MicroMEGAS, made of 5-$\mu$m thick copper, has 35 $\mu$m$^\phi$
holes spaced at intervals of 61 $\mu$m.
The distance between the mesh and the pad plane is maintained to 50 $\mu$m by
kapton pillars arranged in-between.
The typical gain is about 3650 at the mesh potential of -320 V.
The triple GEM, CERN standard, has two 1.5-mm transfer gaps and a 1-mm
induction gap. The transfer and induction fields are 2 kV/cm and
3 kV/cm, respectively.
The typical total effective gain in a P5 (TDR) gas is about 3000
with 335 (340) V applied across each
GEM foil.

The chamber gases are Ar-isobutane (5\%) for MicroMEGAS, and a TDR gas
(Ar-methane (5\%)-carbon dioxide (2\%)) or Ar-Methane (5\%) for GEMs,
at atmospheric pressure and room temperature.
The gas pressure and the ambient temperature are continuously monitored
since they are not controlled actively. 
The drift-field strengths are 200, 220 and 100 V/cm, respectively for
Ar-isobutane, TDR gas and Ar-methane.

The prototype TPC is placed in the uniform field region of a super conducting
solenoid without return yoke, having bore diameter of 850 mm, effective
length of 1000 mm, and the maximum field strength of 1.2 T.
The prototype was then subjected to the beam, mostly 4 GeV/c pions, at the
$\pi$2 test beam facility of the KEK proton synchrotron.

\section{Preliminary results}

In this section we show some preliminary results of the analysis up to now,
only for the data taken with an axial magnetic field of 1 T and with tracks 
normal to the pad rows.
The results of analytic evaluations are used or presented here without 
comments.
Readers are therefore advised to read Appendix and the slides available 
on-line~\cite{LCWSweb} 
as well, where the analytic method is briefly summarized and illustrated.

The observed pad responses for different drift distances ($z$) are shown in
Fig.~2~(a) while the widths of distributions are plotted as a function of
drift distance in Fig.~2~(b).
\begin{figure}[http]
 \hspace{15mm}
  \begin{minipage}{0.8\hsize}
    \begin{tabular}{cc}
      \begin{minipage}[t]{0.48\hsize}
        \centering
        \includegraphics*[scale=0.37]{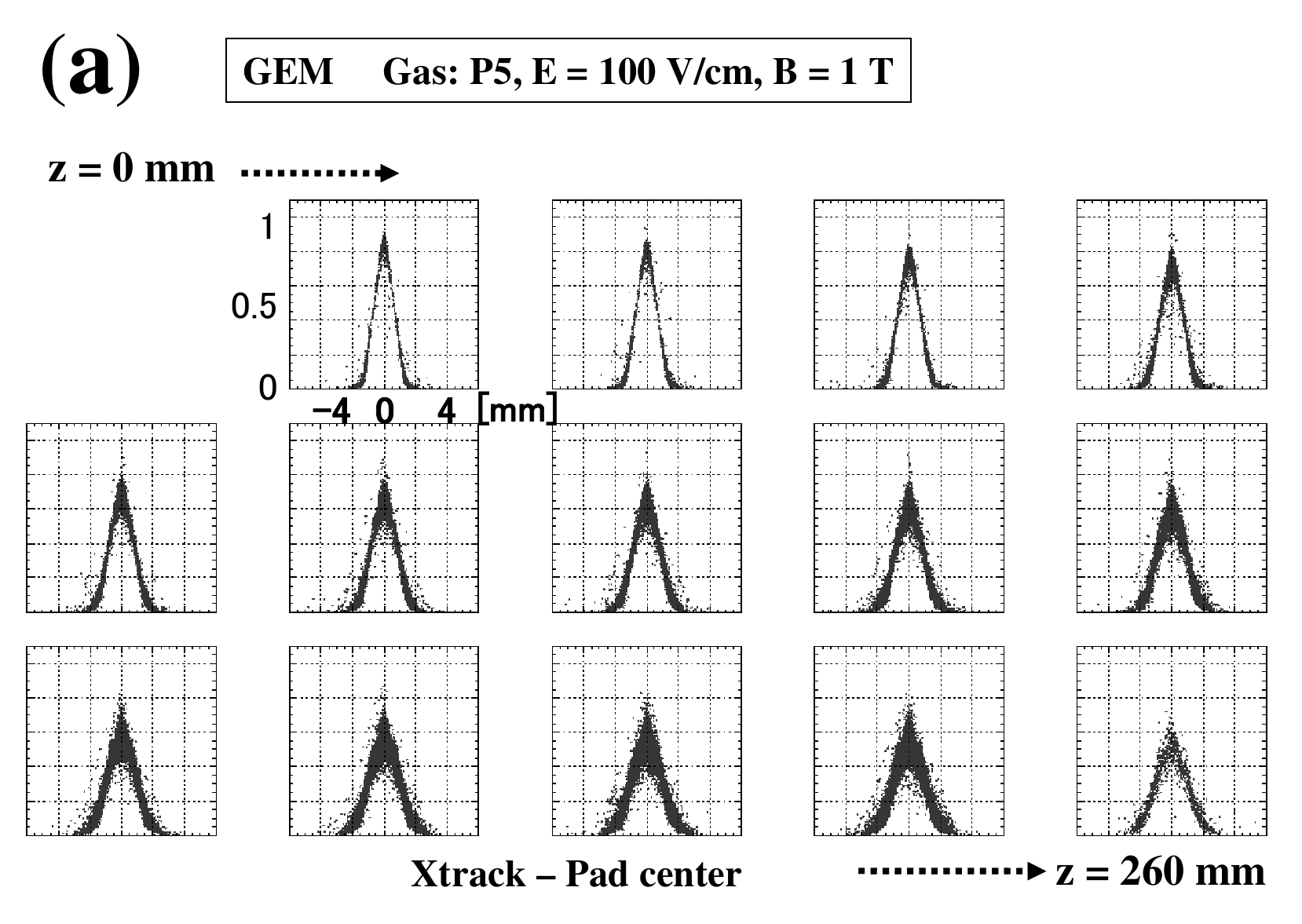}
        \label{fig2a}
      \end{minipage} &
      \begin{minipage}[t]{0.48\hsize}
        \centering
        \includegraphics*[scale=0.37]{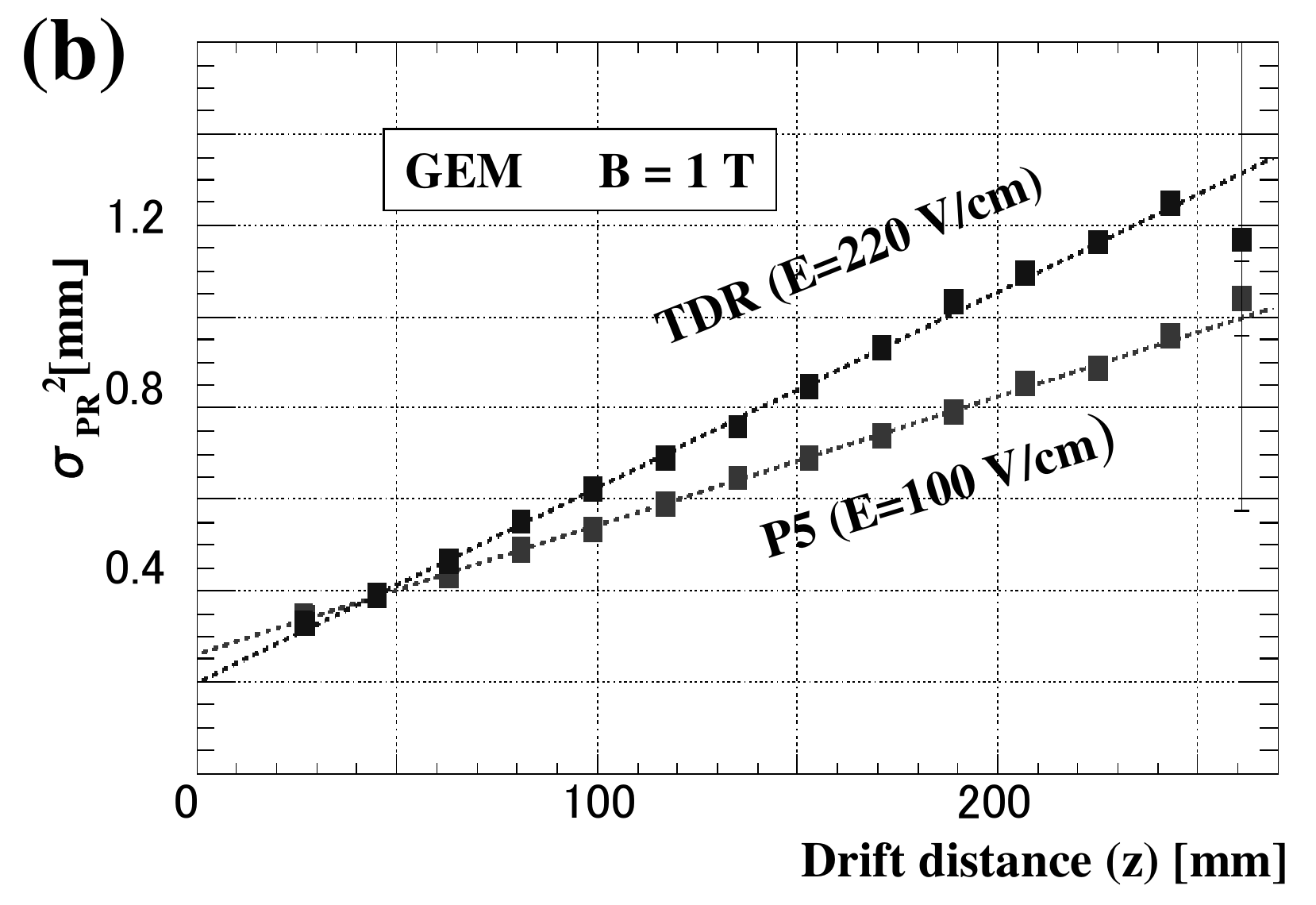}
        \label{fig2b}
      \end{minipage} 
    \end{tabular}
\caption[fig2]{\label{fig2}\footnotesize
(a) Pad responses for different drift distances. (b) Pad-response width
squared ($\sigma_{PR}^2$) vs. drift distance ($z$).
The width of pad response is parametrized as 
$\sigma_{PR}^2 = \sigma_{PR0}^2 + D^2 \cdot z$, 
with $D$ being the diffusion constant.
}
  \end{minipage}
\end{figure}
The measured spatial resolution against drift distance is shown in Fig.~3 (a)
and (b), respectively for the MicroMEGAS and triple GEM readout, along with the
result of the analytic calculation.
In the calculation the pad response function (PRF) was assumed to be $\delta$ 
function for the MicroMEGAS and a Gaussian for the Gems \footnote
{
PRF is the avalanche charge spread on the pad plane for a {\it single\/}
drift electron and should not be confused with the pad response.  
In the case of MicroMEGAS it is much smaller than the pad pitch (2.3 mm)
and is, therefore, neglected.
The width (standard deviation) of the Gaussian PRF for the triple GEM has been
determined from the intercept of the pad-response width squared vs. $z$
(Fig. 2 (b)): $\sigma_{PR}^2 = \sigma_{PR0}^2 + D^2 \cdot z$ with 
$\sigma_{PR0}^2 = w^2 / 12 + \sigma_{PRF}^2$, where the pad pitch $w$ = 1.27 mm
and $\sigma_{PR0} \sim$ 511 $\mu$m, yielding $\sim$ 356 $\mu$m for
$\sigma_{PRF}$.
The value of $\sigma_{PR0}$ thus obtained is consistent with a simple
estimation taking into account only the diffusion in the transfer and
induction gaps. 
}.
\begin{figure}[http]
    \begin{tabular}{cc}
      \vspace*{0.2cm}
      \begin{minipage}[t]{0.48\hsize}
        \centering
        \includegraphics*[scale=0.48]{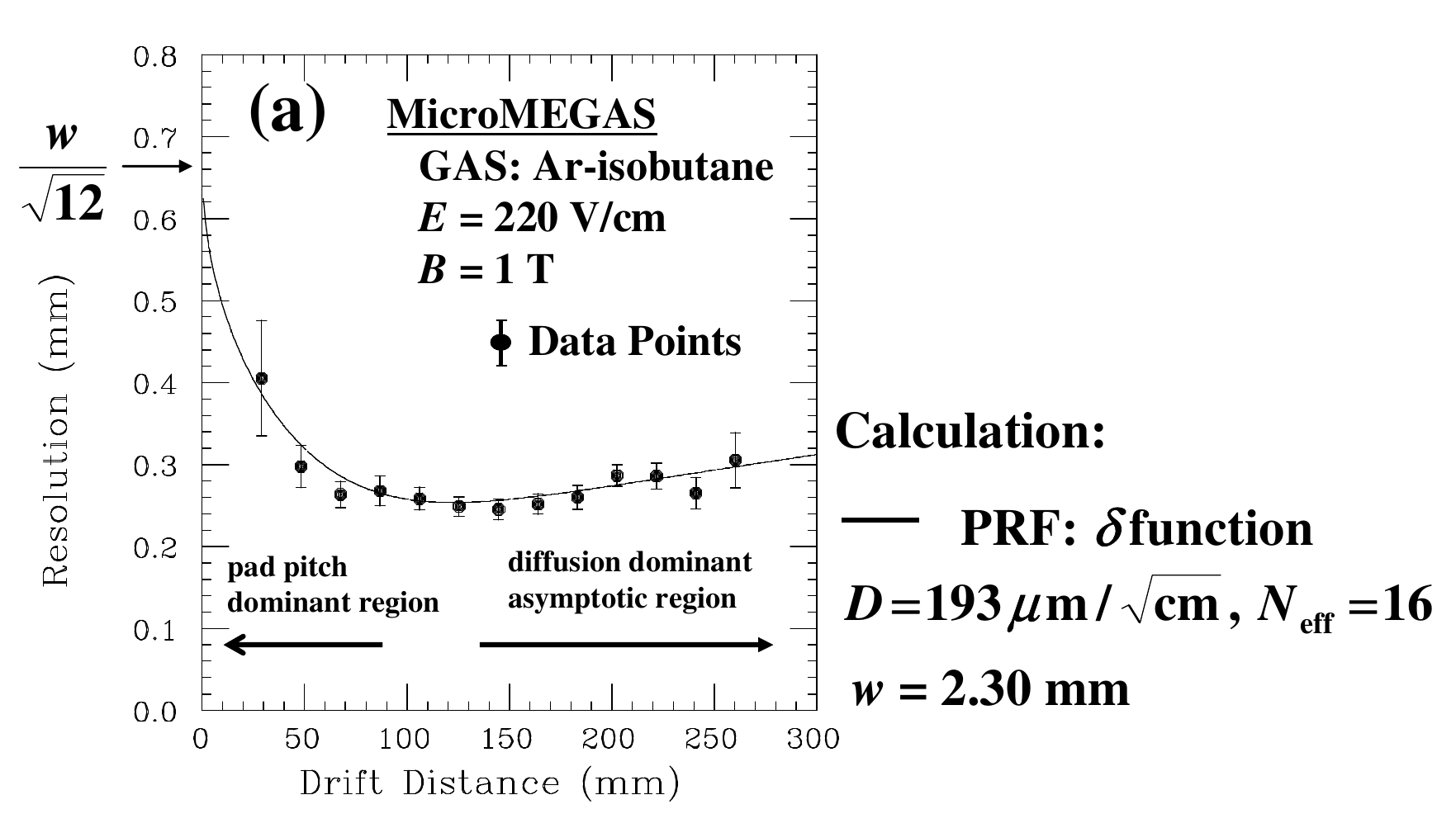}
        \label{fig3a}
      \end{minipage} &
      \vspace*{-0.2cm}
      \begin{minipage}[t]{0.48\hsize}
        \centering
        \includegraphics*[scale=0.48]{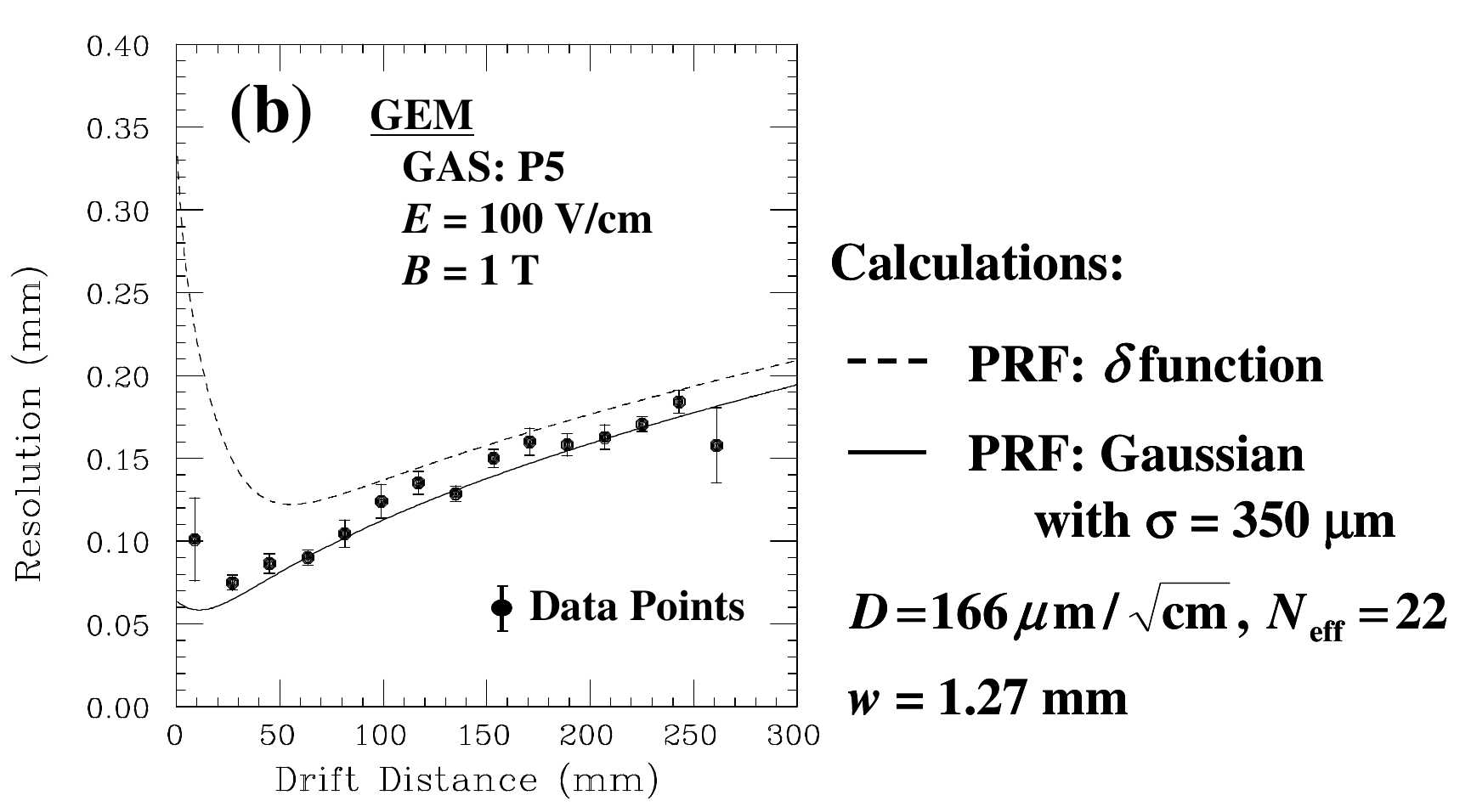}
        \label{fig3b}
      \end{minipage} 
    \end{tabular}
\caption[fig2]{\label{fig3}\footnotesize
(a) Spatial resolution vs. z obtained with MicroMEGAS. Gas: Ar-isobutane (5\%).
(b) Spatial resolution vs. z obtained with GEMs. Gas: Ar-methane (5\%).
}
\end{figure}

The obtained behavior of the pad response, and the spatial resolution
at long drift distances~\footnote
{
When PRF is $\delta$ function the asymptotic behavior of the spatial
resolution at long distances (diffusion dominant asymptotic region)
is described by
$\sigma_X^2 \equiv \sigma_{X0}^2 + D_X^2 \cdot z 
\sim 1./N_{eff} \cdot (w^2/12 + D^2 \cdot z)$, where
$N_{eff}$ is the effective number of electrons and $D$ is
the diffusion constant (see Appendix).
}
are compared with expectations in table 1.
The comparisons show
\begin{enumerate}
\item $\sigma_{PR0}$ is in reasonable agreement with the expectation 
      ($\sqrt{w^2/12 + \sigma_{PRF}^2}$)
      if the contribution of $\sigma_{PRF}$ is taken in to account
      (in the case of GEMs);
\item $\sigma_{X0}$ is in good agreement with the expectation
      ($w/\sqrt{12 \cdot N_{eff}}$) for the MicroMEGAS, and better than this
      for the GEMs because of the significant charge spread in the
      transfer and induction gaps;
\item The values of diffusion constant ($D$) are comparable to those
      given by the simulation (MAGBOLTZ~\cite{Biagi});
\item $N_{eff}$ (16 $\sim$ 22) is significantly smaller than the average number
      of drift electrons per pad row ($\sim$71)~\cite{Kobayashi}.
\end {enumerate}
%
\begin{center}
\begin{minipage}{13cm}

Table 1. Asymptotic behavior at long drift distances under $B$ = 1 T.

\medskip 

(a) Pad response \\
\begin{tabular}{|c||c||c|c|} \hline
{\small Detection device} & MicroMEGAS & \multicolumn{2}{|c|}{GEM} \\
\hline \hline
Gas & {\small Ar-isobutane (5\%)} & \hspace{9mm} TDR \hspace{9mm} &
{\small Ar-methane (5\%)} \\ \hline
$\sigma_{PR0}$ ($\mu$m) & 758 $\pm$ 91 & 432 $\pm$ 3 & 511 $\pm$ 2 \\ \hline
$w/\sqrt{12}$ ($\mu$m) & 664 & \multicolumn{2}{|c|}{367} \\ \hline
$D$ ($\mu$m/$\sqrt{{\rm cm}}$) & 194 $\pm$ 18 & 213 $\pm$ 1 & 168 $\pm$ 1
\\ \hline
$D$ [MAGBOLTZ] & 193 & 200 & 166 \\ \hline
\end{tabular} \\
\medskip \\
(b) Spatial resolution \\
\begin{tabular}{|c||c||c|c|} \hline
{\small Detection device} & MicroMEGAS & \multicolumn{2}{|c|}{GEM} \\
\hline \hline
Gas & {\small Ar-isobutane (5\%)} & \hspace{9mm} TDR \hspace{9mm} &
{\small Ar-methane (5\%)} \\ \hline
$\sigma_{X0}$ ($\mu$m)  & 161 $\pm$ 54 &  44 $\pm$ 10 & 42 $\pm$ 17 \\ \hline
$w/\sqrt{12 \cdot N_{eff}}$ ($\mu$m) & 166 $\pm$ 42 & 86 $\pm$ 3 &
78 $\pm$ 4  \\ \hline 
$D/\sqrt{N_{eff}}$ ($\mu$m/$\sqrt{{\rm cm}}$) & 48 $\pm$ 12 & 47 $\pm$ 1 &
35 $\pm$ 2 \\ \hline
$N_{eff}$ & 16 $\pm$ 8 & 18 $\pm$ 1 & 22 $\pm$ 2 \\ \hline
\end{tabular}  
\end{minipage}
\end{center}
%
%
%

%


\section{Expected spatial resolution of the ILC-TPC}

Calculated spatial resolutions of the ILC-TPC at $B$ = 4 T are shown in
Fig.~4 for tracks perpendicular to the pad row.
\begin{figure}[htbp]
\centering
\includegraphics[width=12.0cm,clip]{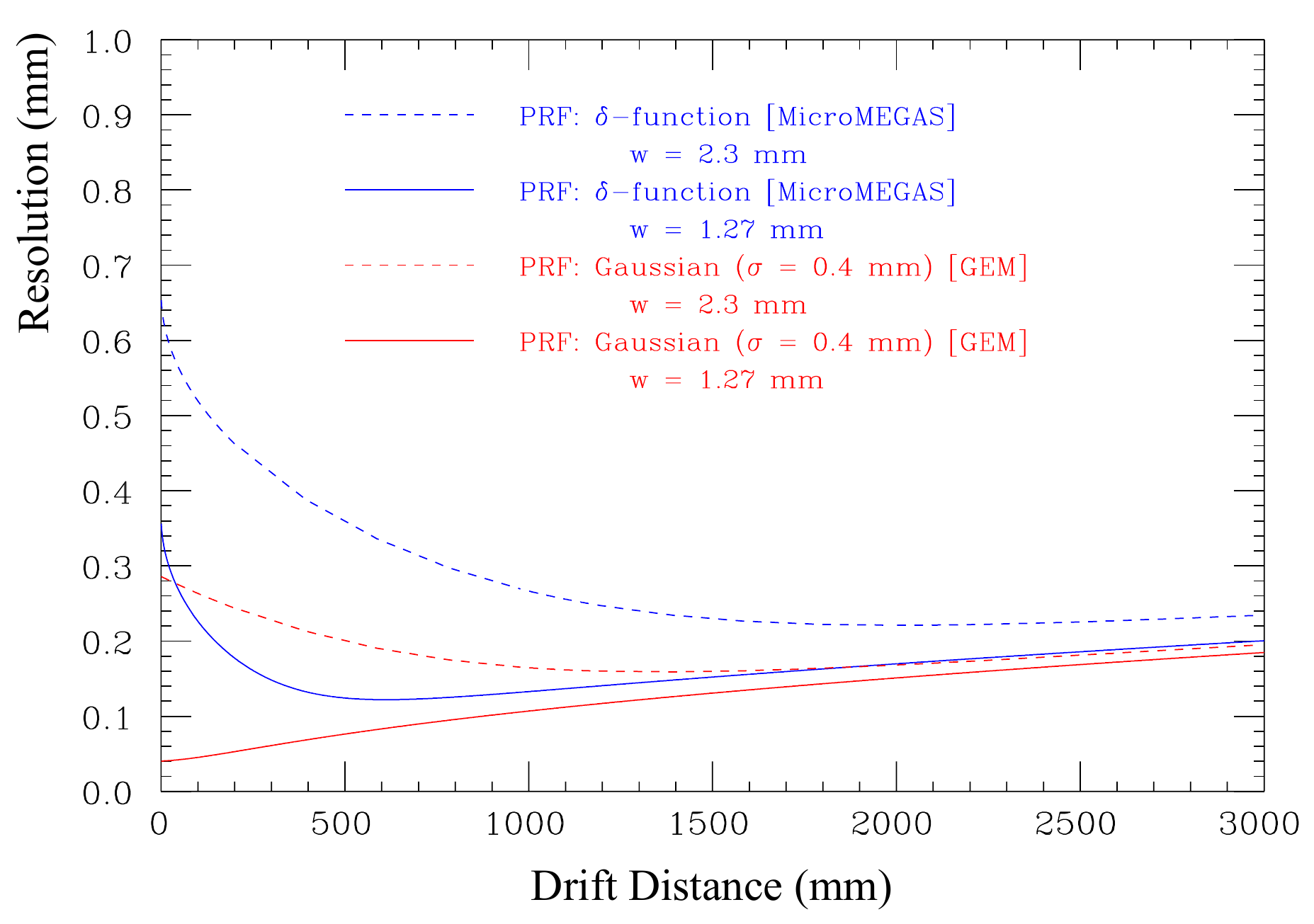}
\caption[fig4]{\label{fig4}\footnotesize
Expected spatial resolutions of the ILC-TPC obtained with MicroMEGAS or GEMs.
Gas: Ar-methane (5\%), $B$ = 4 T ($D$ = 50 $\mu$m/$\sqrt{\rm cm}$),
and $N_{eff}$ = 22.
}
\end{figure}
In the calculations the values of diffusion constants ($D$) given by 
MAGBOLTZ were used.
The figure tells us that under a strong magnetic field it is important to
reduce the pad-pitch dominant region (at small drift distances) in the ILC-TPC
by enhancing the charge sharing among the readout pads, in order to
maintain a good resolution over the entire sensitive volume.

There are several possibilities to realize effective charge sharing:
\begin{itemize}
\item zigzag (chevron) pads.
\item a smaller pad pitch with a larger number of readout channels.
\item defocussing of electrons after gas amplification
      (natural dispersion in the transfer and induction gaps of GEMs,
       {\it stochastic} PRF).
\item Use of resistive anode technique with a moderate number of readout
      channels (applicable to both GEMs and MicroMEGAS,
      {\it static} PRF)~\cite{Dixit}.
\item pixel readout (Digital TPC)~\cite{Colas}.  
\end{itemize}

\section{Summary}

To summarize, 
the prototype TPC equipped with a MicroMEGAS or GEMs operated stably 
during the beam tests.
The tests provided us with valuable information on its performance
under axial magnetic fields of up to 1 T:   
\begin{itemize}
\item The obtained spatial resolution is understood in terms of pad pitch,
      diffusion constant, PRF, and the effective number of electrons.
\item The expected resolution can be estimated by a numerical calculation
      (NOT a Monte-Carlo) for given geometry, gas mixture and PRF if the
      relevant parameters are known.
\item The calculation is based on a simple formula, easy to code and fast,
      though it is applicable only to tracks perpendicular to the pad row.
\item In the case of MicroMEGAS, the spatial resolution as a function of 
      drift distance is well described by the analytic formula,
      assuming $\delta$ function for PRF.
\item In the case of GEMs, the spatial resolution as a function of drift 
      distance is satisfactorily described by the analytic formula, assuming
      a Gaussian for PRF with the width determined from the intercept of the
      pad-response width squared as a function of drift distance.
\item It is important to make the pad pitch small, {\it physically or
      effectively}, in order to reduce both the overall offset term 
      ($\sigma_{X0}$) and the resolution degradation due to finite pad pitch.
\item The spatial resolution required from the ILC-TPC (100 $\sim$ 200 $\mu$m
      for the maximum drift distance of $\sim$ 2.5 m) is now
      within the reach for tracks normal to the pad row.   
\end{itemize}







\appendix

\renewcommand{\thefigure}{A.\arabic{figure}}
\setcounter{figure}{0}

\section*{Appendix: An analytic estimation of pad response
                      and spatial resolution}

%
One way to estimate the spatial resolution of a TPC is to write a
realistic Monte-Carlo simulation code.
This technique is applicable to any situation, and has been developed by several
groups.
On the other hand, an analytic approach is applicable only to a restricted
case where incident particles are normal to the pad row.
However, the resultant formula is rather simple and is sometimes
enlightening as shown below.
Though a numerical calculation is needed to evaluate the formula, the demanded
CPU time is much less than a Monte-Carlo simulation.
In addition, the analytic calculation can be used to check the reliability of a
Monte-Carlo simulation program, which is usually long and complicated.
This appendix is devoted to briefly summarize our analytic approach,
based on the following assumptions:
\begin{enumerate}
\item Particle tracks are normal to the pad row;
\item Track coordinate is determined by the charge centroid method;
\item Contribution of ambient electronic noise is negligible;
\item Displacement of arriving drift electrons due to $E \times B$ effect
      near the entrance to the detection device is negligible;
\item Displacement of arriving electrons due to the finite granularity of
      amplification elements of the detection device
      (line intervals in MicroMEGAS or a hole pitch in GEM)
      is negligible.
\end{enumerate}
\renewcommand{\thesection}{\Alph{subsection}}
\subsection{Pad response}

\begin{figure}[htbp]
\centering
\includegraphics[width=10.0cm,clip]{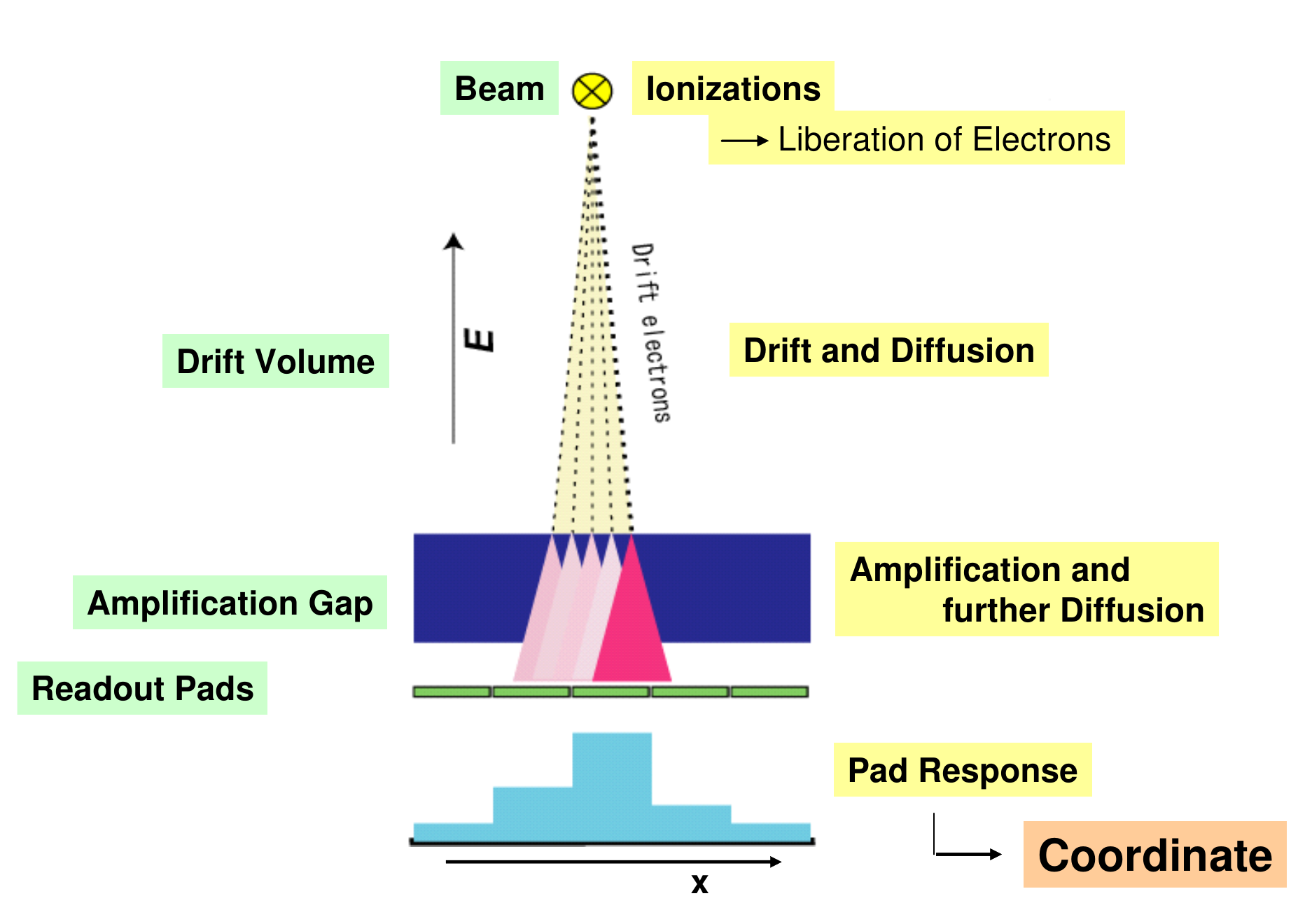}
\caption[figA1]{\label{figA1}
\footnotesize Principle of the track coordinate measurement.
}
\end{figure}

\begin{figure}[htbp]
\centering
\includegraphics[width=12.0cm,clip]{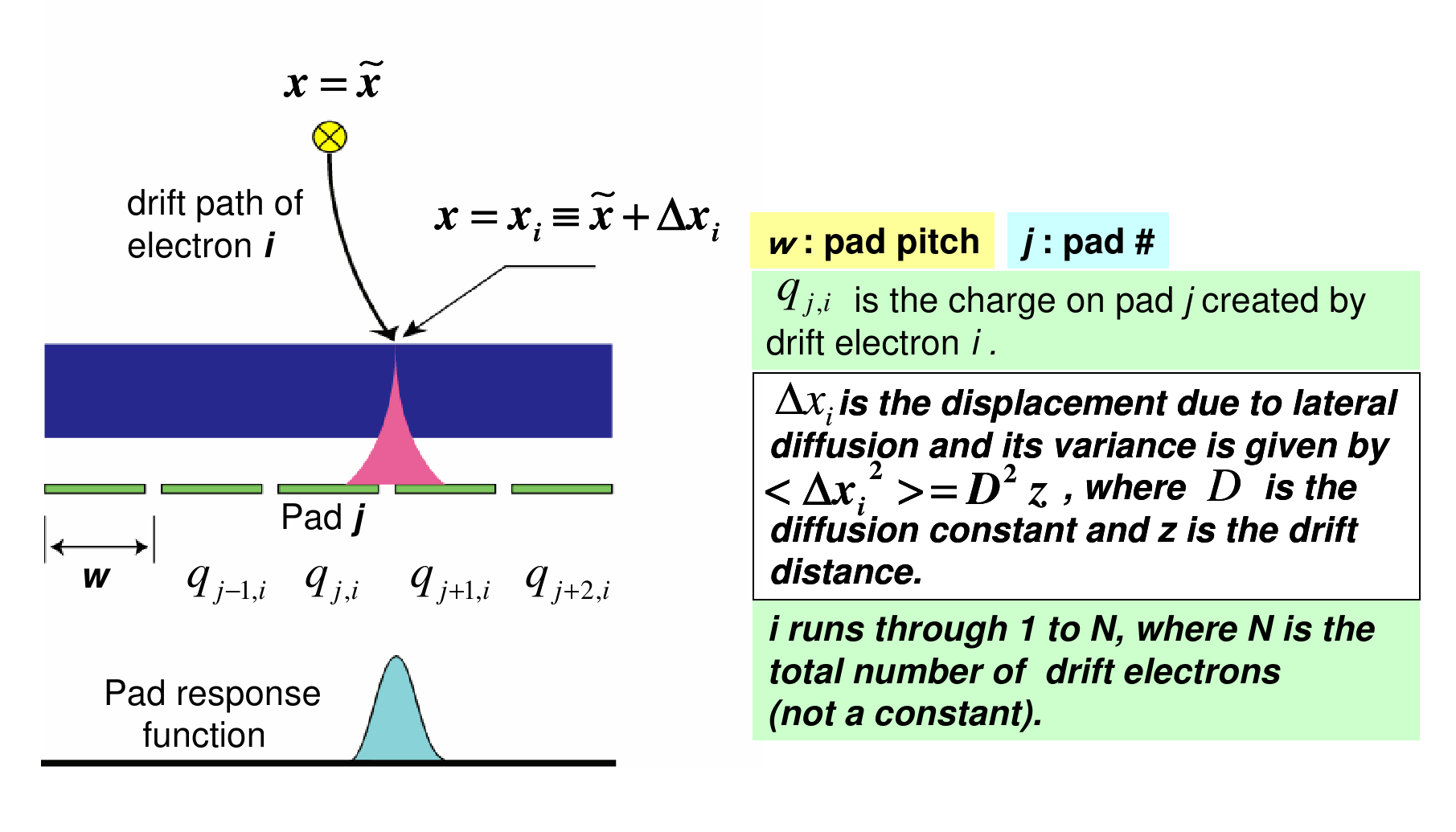}
\caption[figA2]{\label{figA2}
\footnotesize Illustration of the relevant parameters.
}
\end{figure}

Let us calculate here the width of pad response with respect to
the true coordinate assuming that the "pad response function
(PRF)~\footnote
{
In the case of conventional MWPC readout, PRF is defined as the charge
distribution on the pad plane caused by a single drift electron arriving at
a sense wire.
Therefore it is {\it static} and is determined electro-statically.
On the other hand, in the case of MicroMEGAS or GEMs the charge distribution
for a single drift electron is caused mainly by avalanche spread due to
diffusion or by diffusion in the transfer and induction gaps.
Therefore it is essentially {\it stochastic\/}.
In the analytic approach discussed here, however, PRF is treated as if
it were {\it static\/}, assuming a large avalanche multiplication factor.
}"
is $\delta$ function. 

\begin{displaymath}
\left< (x^\# - \tilde{x})^2 \right> = \sum_{N=1}^{\infty} P(N) \cdot
\frac{1}{w} \cdot \int_{-w/2}^{+w/2} d\tilde{x}\;
\Bigl( \prod_{k=1}^N \int P_x (x_k) dx_k \int P_q(q_k)dq_k \Bigr)
\sum_{i=1}^N 
\frac {q_i}{\sum_{j=1}^N q_j} \cdot (x_i^\# - \tilde{x})^2\;,
\end{displaymath}
where $P(N)$ is the probability density function (PDF) of total number of drift
electrons ($N$), $w$ is the pad pitch, $P_x(x_k)$ is the PDF of $k$-th electron's
arrival position ($x_k$), $P_q(q_k)$ is the PDF of $k$-th electron's
signal charge ($q_k$), $x_i^\#$ is the central coordinate of
the pad on which $i$-th electron arrives
($= j \cdot w$, with $j$ being the corresponding pad number),
and $\tilde{x}$ is the original position (true coordinate) of electrons.    
$P_x(x)$, accounting for diffusion, is denoted later by
$P_x(x;\tilde{x},\sigma_d)$, where $\tilde{x}$ ($\sigma_d$) is
the mean (width) of a Gaussian distribution:
\begin{displaymath}
P_x (x) \equiv P_x(x;\tilde{x},\sigma_d)
        \equiv \frac{1}{\sqrt{2\pi}\sigma_d} \cdot {\rm exp} \left(
                    - \frac{(x-\tilde{x})^2}{2\sigma_d^2} \right) \;.
\end{displaymath}
Figs.~\ref{figA1} and \ref{figA2} illustrate the situation and give some of the
definitions of relevant variables.

\vspace*{0mm}


The calculation proceeds straightforwardly as follows:

\begin{eqnarray*}
\left< (x^\# - \tilde{x})^2 \right> &=& \sum_{N=1}^{\infty} P(N) \cdot
    \sum_{i=1}^N \Bigl( \prod_{k=1}^N \int P_q(q_k)dq_k \Bigr)
    \sum_{i=1}^N \frac{q_i}{\sum_{j=1}^N q_j} \cr
&& ~~~~~~~~~~~~~~~~~~~~~~~~~~~
    \times \frac{1}{w} \cdot \int_{-w/2}^{+w/2} d\tilde{x} \,
    \Bigl( \prod_{k=1}^N \int_{\infty}^{\infty} P_x (x_k) dx_k \Bigr )
    \cdot (x_i^\# - \tilde{x})^2 \\
&=& \sum_{N=1}^{\infty} P(N) \cdot \frac{1}{N} \cdot \sum_{i=1}^{N}
    \frac{1}{w} \cdot \int_{-w/2}^{+w/2} d\tilde{x} \, 
    \Bigl( \prod_{k=1}^{N} \int_{-\infty}^{\infty} P_x(x_k)dx_k \Bigr)
    \cdot (x_i^\# - \tilde{x})^2 \\
&=& \sum_{N=1}^{\infty} P(N) \cdot \frac{1}{w}
        \cdot \int_{-w/2}^{+w/2} d\tilde{x} \int_{-\infty}^{\infty} P_x(x)
                             \cdot (x^\# - \tilde{x})^2 dx \\
&=& \frac{1}{w} \cdot \int_{-w/2}^{+w/2} d\tilde{x}
          \int_{-\infty}^{\infty} P_x(x) \cdot
                                       (x^\# - \tilde{x})^2 dx \\
&=& \frac{1}{w} \cdot \int_{-w/2}^{+w/2} d\tilde{x}
                 \sum_{j=-\infty}^{\infty} \int_{jw-w/2}^{jw+w/2} 
                         P_x(x;\tilde{x},\sigma_d) \cdot (jw-\tilde{x})^2 dx \\ 
&=& \frac{1}{w} \cdot \sum_{j=-\infty}^{\infty}
                 \int_{-w/2}^{+w/2} d\tilde{x} (jw-\tilde{x})^2 
                   \int_{jw-w/2}^{jw+w/2} P_x(x;\tilde{x},\sigma_d) dx \\
&=& \frac{1}{w} \cdot \sum_{j=-\infty}^{\infty} \int_{jw-w/2}^{jw+w/2}dt \cdot
                         t^2 \int_{jw-w/2}^{jw+w/2} P_x(x;jw-t,\sigma_d) dx
~~~~~~                                {\rm with}~t \equiv jw - \tilde{x} \\
&=& \frac{1}{w} \cdot \sum_{j=-\infty}^{\infty} \int_{jw-w/2}^{jw+w/2}dt \cdot
                         t^2 \int_{-w/2}^{+w/2} P_x(x;-t,\sigma_d) dx \\
&=& \frac{1}{w} \cdot \int_{-\infty}^{\infty} dt \cdot t^2
                     \int_{-w/2}^{+w/2} P_x(x;-t,\sigma_d)dx \\
&=& \frac{1}{w} \cdot \int_{-w/2}^{+w/2} dx 
                     \int_{-\infty}^{\infty} t^2 \cdot P_x(x;-t,\sigma_d)dt \\
&=& \frac{1}{w} \cdot \int_{-w/2}^{+w/2} dx 
             \int_{-\infty}^{\infty} (u-x)^2 \cdot P_x(u;0,\sigma_d)du
~~~~~~~{\rm with}~u \equiv x + t \\
&=& \frac{1}{w} \cdot \int_{-w/2}^{+w/2} (\sigma_d^2 + x^2) \; dx \\
&=& \sigma_d^2 + \frac{w^2}{12} \; .
~~~~~~~~~~~~~~~~~~~~~~~~~~~~~~~~~~~~~~~~~~~~~~~~~~~~~~~~~~~~~~~~~~~~~~~~~~(A.1)
\end{eqnarray*}

The interpretation of the result is quite simple.
The squared pad-response width is a quadratic sum of the widths, one due to
diffusion and the other originated from the finite pad pitch.
This can be readily generalized for the case where the width of PRF 
($\sigma_{PRF}$) is finite:
\begin{displaymath}
\left< (x^\# - \tilde{x})^2 \right> = \sigma_d^2 + \sigma_{PRF}^2 
                         + \frac{w^2}{12}
           = \frac{w^2}{12} + \sigma_{PRF}^2 + D^2 \cdot z \; ,
~~~~~~~~~~~~~~~~~~~~~~~~~~~~~~~~~~~(A.2)
\end{displaymath}
where $D$ is the diffusion constant and $z$ is the drift distance.
Therefore if the square of the width of pad response is plotted against $z$
one gets a straight line with a slope of $D^2$ and an intercept of
$w^2/12 + \sigma_{PRF}^2$.

In fact, we use the width of pad response with respect to the charge centroid
($\equiv \bar{x}$), instead of the unknown (precise) true coordinate
($\tilde{x}$), in the present paper.
Therefore Eq. (A.1) needs a slight modification accordingly
as briefly shown below for the case where PRF is $\delta$ function
($\sigma_{PRF} = 0$).
In the calculation, signal charge fluctuation represented by 
$P_q(q)$ is not included explicitly since it does not affect the final result. 
From now on we avoid to explicitly show the integrals weighted by 
PDFs and use instead average symbols
denoted by $\left< \cdot \cdot \cdot \cdot \cdot \right>$ 
in order to save space. 
\begin{eqnarray*}
\left< (x^\# - \bar{x})^2 \right>
&=& \frac{1}{N} \cdot \left< \sum_{i=1}^{N} (x_i^\# - \bar{x})^2 \right > \\
&=& \frac{1}{N} \cdot \left< \sum_{i=1}^{N} \left( (x_i^\#-\tilde{x})
                                              - (\bar{x}-\tilde{x}) \right)^2 \right> \\ 
&=& \frac{1}{N} \cdot \left< \sum_{i=1}^N \left( (x_i^\# - \tilde{x})^2
                             + (\bar{x} - \tilde{x})^2
          -2 \cdot (x_i^\# - \tilde{x}) \cdot (\bar{x} - \tilde{x}) \right) \right> \\
&=& \left< (x^\# - \tilde{x})^2 \right> + \left< (\bar{x} - \tilde{x})^2 \right>
             - \frac{2}{N} \cdot \left< (\bar{x} - \tilde{x}) \cdot 
                                     \sum_{i=1}^{N} (x_i^\# - \tilde{x}) \right> \\
&=& \left< (x^\# - \tilde{x})^2 \right> + \left< (\bar{x} - \tilde{x})^2 \right>
              - 2 \cdot \left< ( \bar{x} - \tilde{x} )^2 \right> \\
&=& \left< (x^\# - \tilde{x})^2 \right> - \left< (\bar{x} - \tilde{x})^2 \right> \;.
~~~~~~~~~~~~~~~~~~~~~~~~~~~~~~~~~~~~~~~~~~~~~~~~(A.3)
\end{eqnarray*}

The first term is what we have calculated above (Eq. (A.1)) while the second term
is nothing but the spatial resolution (squared) obtained with
the charge centroid method, which is to be evaluated in the next section. 
The contribution of second term is small except at small drift distances.

\renewcommand{\thesection}{\Alph{section}}
\addtocounter{section}{1}
\subsection{Spatial resolution}

Let us consider first the spatial resolution to be obtained with infinitesimal
pad pitch and $\sigma_{PRF}$ (PRF: $\delta$ function) since the calculation
is very simple in this case~\cite{Kobayashi}.
In the following, the measured track coordinate is assumed to be determined by
the centroid of charges collected by the readout pads:
\begin{displaymath}
X \equiv \frac{\sum_{i=1}^{N} q_i \cdot x_i}{\sum_{i=1}^{N} q_i} \; ,
\end{displaymath}
where $q_i$, $x_i$ are the signal charge and the arrival position,
respectively, of $i$-th electron.  
In the calculation below and in the rest of this appendix, the symbol
$<.....>_{x~(q)}$ stands for the average taken over the variables $x~(q)$
with the corresponding PDFs.
The subscript $x$ or $q$ may be omitted when the meaning of average is
clear itself. Then 
\begin{eqnarray*}
\sigma_X^2 &=& \Bigl<(X-\tilde{x})^2 \bigr > \\
&=& \biggl< \Bigl(
        \frac{\sum_{i=1}^N q_i \cdot (x_i-\tilde{x})}{\sum_{i=1}^N q_i}
                                                           \Bigr)^2 \biggr> \\
&=& \biggl< \frac{1}{(\sum_i q_i)^2} \cdot
            \biggl( \sum_{i} q_i^2 \cdot (x_i-\tilde{x})^2
               + \sum_{i \neq j} q_i \cdot q_j
                      \cdot (x_i-\tilde{x}) \cdot (x_j-\tilde{x})
                                                            \biggr)
                                                                 \biggr> \\
&=& \biggl<
       \frac{1}{(\sum_i q_i)^2} \cdot
            \biggl( \Bigl< (x-\tilde{x})^2 \Bigr>_x \cdot \sum_{i} q_i^2 
          + \bigl< x-\tilde{x} \bigr>_x^2 \cdot \sum_{i \neq j} q_i \cdot q_j
                                                            \biggr)
                                                                \biggr>_q \\
&=& \biggl< \frac{\sum_{i} q_i^2 }{(\sum_i q_i)^2} \biggr>_q
                \cdot \Bigl< (x-\tilde{x})^2 \Bigr>_x \\
&=& \biggl< \frac{\sum_{i} q_i^2 }{(\sum_i q_i)^2} \biggr>_q \cdot \sigma_d^2 \\
&\approx& \frac{1}{N} \cdot \frac{\bigl< q^2 \bigr>}{\bigl< q \bigr>^2}
                                                 \cdot \sigma_d^2 \; ,
~{\rm assuming}~\sum_i q_i = {\rm const} = N \cdot \bigl< q \bigr> \;,
~{\rm expecting~a~large}~N. \\
{\rm Averaging~over}~N \; , {\rm ~we~obtain} \hspace{-15mm} \\
%
\sigma_X^2 &\approx& \sum_{i=1}^{\infty} P(N) \cdot \frac{1}{N}
          \cdot \frac{\bigl< q^2 \bigr>}{\bigl< q \bigr>^2} \cdot \sigma_d^2 \\
&\approx& \biggl< \frac{1}{N} \biggr> 
          \cdot \frac{\bigl< q^2 \bigr>}{\bigl< q \bigr>^2} \cdot \sigma_d^2 \\
&\approx& \frac{1}{N_{eff}} \cdot \sigma_d^2 \; ,
~~~~~~~~~~~~~~~~~~~~~~~~~~~~~~~~~~~~~~~~~~~~~~~~~~~~~~~~~~~~~~~~~~(A.4)
\end{eqnarray*}
where $N_{eff}$ is defined as
\begin{displaymath}
\frac{1}{N_{eff}} \equiv \biggl< \frac{1}{N} \biggr>
     \cdot \frac{\bigl< q^2 \bigr>}{\bigl< q \bigr>^2}
           \equiv \biggl< \frac{1}{N} \biggr> \cdot (1+K) \;,~~~~~~~~~~~~~~~~~~
\end{displaymath}
with $K$ being the relative variance of avalanche fluctuation:
$\sigma_q^2 / \bigl< q \bigr>^2 \;$.

Next, let us assume a finite pad pitch ($w$) but still an infinitesimal
PRF width ($\sigma_{PRF}$).
In this case, the charge centroid is given by
\begin{displaymath}
X = \frac{\sum_{i=1}^N q_i \cdot x_i^\#}{\sum_{i=1}^{N} q_i} \; ,
\end{displaymath}
where $x_i^\#~(= j \cdot w$) is the central coordinate of the pad on which 
electron $i$ arrives, and  
\begin{eqnarray*}
\sigma_X^2 &=& \Bigl<(X-\tilde{x})^2 \bigr > \\
&=& \biggl< \Bigl(
        \frac{\sum_{i=1}^N q_i \cdot (x_i^\#-\tilde{x})}{\sum_{i=1}^N q_i}
                                                           \Bigr)^2 \biggr> \\
&=& \biggl< \frac{1}{(\sum_i q_i)^2} \cdot
            \biggl( \sum_{i} q_i^2 \cdot (x_i^\#-\tilde{x})^2
               + \sum_{i \neq j} q_i \cdot q_j
                      \cdot (x_i^\#-\tilde{x}) \cdot (x_j^\#-\tilde{x})
                                                            \biggr)
                                                                 \biggr> \\
&=& \biggl<
       \frac{1}{(\sum_i q_i)^2} \cdot
            \biggl( \Bigl< (x^\#-\tilde{x})^2 \Bigr>_x \cdot \sum_{i} q_i^2 
              + \bigl< x^\#-\tilde{x} \bigr>_x^2 
                \cdot \biggl( \sum_{i,j} q_i \cdot q_j - \sum_i q_i^2 \biggr)
                                                             \biggr)
                                                                \biggr>_q \\
&=& \bigl< x^\#-\tilde{x} \bigr>^2 + 
    \biggl< \frac{\sum_{i} q_i^2 }{(\sum_i q_i)^2} \biggr>
         \cdot \biggl(\bigl< (x^\#-\tilde{x})^2 \bigr>
                               - \bigl< x^\#-\tilde{x} \bigr>^2 \biggr) \\
&\approx& \bigl<x^\# - \tilde{x}\bigr>^2 + 
     \frac{1}{N} \cdot \frac{\bigl<q^2\bigr>}{\bigl<q\bigr>^2}
       \cdot \bigl( \bigl<(x^\#)^2\bigr> - \bigl<x^\#\bigr>^2 \bigr) \;.\\
\end{eqnarray*}
Averaging over $N$, and substituting $j \cdot w$ for $x^\#$, we obtain
\begin{eqnarray*}
\sigma_X^2 &\approx& \bigl<x^\# - \tilde{x}\bigr>^2 + 
         \frac{1}{N_{eff}}
              \cdot \bigl( \bigl<(x^\#)^2\bigr> - \bigl<x^\#\bigr>^2 \bigr) \\
&\approx& \biggl( \sum_{j=-\infty}^{\infty} jw \cdot P^\#_x(jw)
                                                 -\tilde{x} \biggr)^2
     + \frac{1}{N_{eff}} \cdot \biggl(
         \sum_{j=-\infty}^{\infty} j^2w^2 \cdot P^\#_x(jw)  
          - \biggl( \sum_{j=-\infty}^{\infty} jw \cdot P^\#_x(jw) \biggr)^2 \;
                                                         \biggr) \;,
\end{eqnarray*}
\begin{displaymath}
\hspace{-13mm}
{\rm where~~} P_x^\# (jw) \equiv \int_{jw-w/2}^{jw+w/2} P_x (x) dx,
{\rm ~with~} P_x (x) \equiv \frac{1}{\sqrt{2\pi}\sigma_d}
          {\rm exp} \biggl( -\frac{(x-\tilde{x})^2}{2\sigma_d^2} \biggr)\; .
~~~~~~~~~~~~~~~~~~~~~(A.5)
\end{displaymath}

The first term in the final expression originates from the bias due to the
charge centroid method combined with the finite pad pitch.
This term is independent of $N$ and rapidly decreases with increasing $z$
because of diffusion~\cite{LCWSweb}.
On the other hand,  the second term is the square of 
{\it observed} charge spread relative to the charge centroid
(Eq. (A.3): $\sim \sigma_d^2 + w^2/12$) divided by $N_{eff}$.

Finally let us assume a finite PRF width ($\sigma_{PRF}$).
In this case, the charge centroid is given by
\begin{displaymath}
X = \frac{\sum_{i=1}^N \sum_j q_{ji} \cdot x_j^*}
                        {\sum_{i=1}^{N} \sum_j q_{ji}}
\equiv \frac{\sum_{i=1}^N Q_i \sum_j F_j(x_i) \cdot x_j^*}{\sum_{i=1}^N Q_i}\;,
\end{displaymath}
where (see Fig.~\ref{figA2})
\begin{eqnarray*}
q_{ji} &\equiv& Q_i \cdot F_j (x_i) {\rm~:~signal~charge~on~pad~} j,
                    {\rm ~created~by~electron~} i \;, \\
x_i &:& {\rm arrival~position~of~electron~} i {\rm ~at~the~entrance~to
                                              ~the~detection~device} \;, \\  
x_j^* &\equiv& j \cdot w {\rm ~~:~central~coordinate~of~pad~} j \;\;\;
      (j = \cdot\cdot\cdot, -2,-1,0,+1,+2, \cdot\cdot\cdot) \;, \\
Q_i &\equiv& \sum_j q_{ji}
            {\rm ~:~total~signal~charge~created~by~electron~} i \;, \\ 
F_j(x_i) &\equiv& \frac{q_{ji}}{Q_i}
             \equiv \int_{jw-w/2}^{jw+w/2} f(\xi-x_i) d\xi \;, \\
f(\xi) &:& {\rm (normalized)~PRF} \; . 
\end{eqnarray*}
And
\begin{eqnarray*}
\sigma_X^2 &=& \Bigl<(X-\tilde{x})^2 \bigr > \\
&=& \biggl< \biggl(
        \frac{\sum_{i=1}^N Q_i \sum_j F_j(x_i) \cdot x_j^*}{\sum_{i=1}^N Q_i}
              - \tilde{x} \biggr)^2 \biggr> \\
&=& \biggl< \biggl(
        \frac{\sum_{i=1}^N Q_i \sum_j F_j(x_i) \cdot (x_j^*-\tilde{x})}
                                           {\sum_{i=1}^N Q_i}
                                                \biggr)^2 \biggr> \\
&=& \biggl< 
     \frac{1}{(\sum_i Q_i)^2} \cdot 
       \biggl(
         \sum_i Q_i^2 \cdot
           \biggl(
\sum_j F_j(x_i) \cdot (x_j^*-\tilde{x}) \sum_k F_k(x_i) \cdot (x_k^*-\tilde{x})
           \biggr) \\ 
&& ~~~~~~~~~~~~~~~~~~~~+
\sum_{i\neq j} Q_iQ_j \cdot
           \biggl(
\sum_k F_k(x_i) \cdot (x_k^*-\tilde{x}) \sum_l F_l(x_j) \cdot (x_l^*-\tilde{x})
           \biggr) 
                   \biggr) \biggr > \\
&=& \biggl< 
     \frac{1}{(\sum_i Q_i)^2} \cdot 
       \biggl(
         \sum_i Q_i^2 \cdot
           \biggl(
\Bigl( \sum_j F_j(x_i) \cdot x_j^*-\tilde{x} \Bigr)
                  \Bigl( \sum_k F_k(x_i) \cdot x_k^*-\tilde{x} \Bigr)
           \biggr) \\ 
&& ~~~~~~~~~~~~~~~~~~~~+
\sum_{i\neq j} Q_iQ_j \cdot
           \biggl(
\Bigl( \sum_k F_k(x_i) \cdot x_k^*-\tilde{x} \Bigr)
                  \Bigl( \sum_l F_l(x_j) \cdot x_l^*-\tilde{x} \Bigr)
           \biggr) 
                   \biggr) \biggr > \\
&=& \biggl< 
     \frac{1}{(\sum_i Q_i)^2} \cdot 
       \biggl(
         \sum_i Q_i^2 \cdot
           \biggl<
\Bigl( \sum_j F_j(x) \cdot x_j^*-\tilde{x} \Bigr)
                  \Bigl( \sum_k F_k(x) \cdot x_k^*-\tilde{x} \Bigr)
           \biggr>_x \\ 
&& ~~~~~~~~~~~~~~~~~~~~~~~~~~~~~~~~~~~ +
\bigl( \sum_{i,j} Q_iQ_j - \sum_i Q_i^2 \bigr) \cdot
           \biggl<
    \sum_k F_k(x) \cdot x_k^*-\tilde{x}
           \biggr>_x^2~ 
                   \biggr) \biggr >_q \\
&=& \biggl< \sum_j F_j(x) \cdot x_j^*-\tilde{x} \biggr>_x^2 
+ \biggl< \frac{\sum_i Q_i^2}{(\sum_i Q_i)^2} \biggr> \cdot
\biggl( \biggl< \sum_{j,k} F_j(x) \cdot F_k(x) \cdot x_j^* \cdot x_k^* \biggr>_x
- \biggl< \sum_k F_k(x) \cdot x_k^* \biggr>_x^2~ \biggr) \\
&=& \biggl( \sum_j \bigl< F_j(x) \bigr> \cdot x_j^*-\tilde{x} \biggr)^2 
+ \biggl< \frac{\sum_i Q_i^2}{(\sum_i Q_i)^2} \biggr> \cdot
\biggl(  \sum_{j,k} \bigl< F_j(x) \cdot F_k(x) \bigr> \cdot x_j^* \cdot x_k^*
- \biggl(  \sum_k \bigl< F_k(x) \bigr> \cdot x_k^* \biggr)^2~ \biggr) \\
&\approx& \biggl( \sum_j \bigl< F_j(x) \bigr> \cdot x_j^*-\tilde{x} \biggr)^2 
+ \frac{1}{N} \cdot \frac{\bigl< Q^2 \bigr>}{\bigl< Q \bigr>^2} \cdot
\biggl(  \sum_{j,k} \bigl< F_j(x) \cdot F_k(x) \bigr> \cdot x_j^* \cdot x_k^*
- \biggl(  \sum_k \bigl< F_k(x) \bigr> \cdot x_k^* \biggr)^2~ \biggr)\;. \\
&& \hspace{-22mm} {\rm Averaging~over~} N\;, {\rm~and~substituting~} j \cdot w {\rm ~and~}
k \cdot w, {\rm ~respectively~ for~} x_j^* ~and~ x_k^* \;,{\rm ~we~get} \\ 
\sigma_X^2 &\approx& \biggl( \sum_j jw \cdot \bigl< F_j(x) \bigr>-\tilde{x} \biggr)^2 
+ \frac{1}{N_{eff}} \cdot
\biggl(  \sum_{j,k} jkw^2 \cdot \bigl< F_j(x) \cdot F_k(x) \bigr> 
- \biggl(  \sum_j jw \cdot \bigl< F_j(x) \bigr> \biggr)^2~ \biggr), ~(A.6) \\
%
{\rm where} && \\
&& \bigl< F_j(x) \bigr > \equiv \int_{-\infty}^{\infty} P_x(x) \cdot F_j(x) \; dx \;,\\
&& \bigl< F_j(x) \cdot F_k(x) \bigr > \equiv
             \int_{-\infty}^{\infty} P_x(x) \cdot F_j(x) \cdot F_k(x) \; dx \;,\\
&& {\rm with} \\ 
&& P_x(x) \equiv \frac{1}{\sqrt{2\pi} \sigma_d}
                {\rm exp} \biggl( - \frac{(x-\tilde{x})^2}{2\sigma_d^2} \biggr) \;,\\
&& F_j(x) \equiv \int_{jw-w/2}^{jw+w/2} f(\xi-x) d\xi \;, \\
&& f(\xi) : {\rm ~PRF} \; .
\end{eqnarray*}
%

It should be pointed out here that $\sigma_X^2$ depends on the position of
$\tilde{x}$ relative to the corresponding pad center, and that
the beam spot size is usually much larger than the pad pitch.
Therefore unless the incident positions of incoming particles are measured
precisely by an external tracker (e.g. by a set of silicon strip detectors)
on an event-by-event basis, $\sigma_X^2$ obtained above (Eq. (A.5) or (A.6))
has to be averaged over $\tilde{x}$ in a range, say, [-$w$/2, +$w$/2].

It is easy to show that Eq. (A.6) is a generalization of Eqs. (A.4) and (A.5).
Eq. (A.5) is expected to be a good approximation when $\sigma_{PRF}$ is 
much smaller than the pad pitch $w$, i.e. in the case of MicroMEGAS.
On the other hand, Eq. (A.6) has to be used for GEM readout since 
$\sigma_{PRF}$ is several hundred microns and is not negligible 
as compared to $w$.

\begin{figure}[htbp]
\centering
\includegraphics[width=14cm,clip]{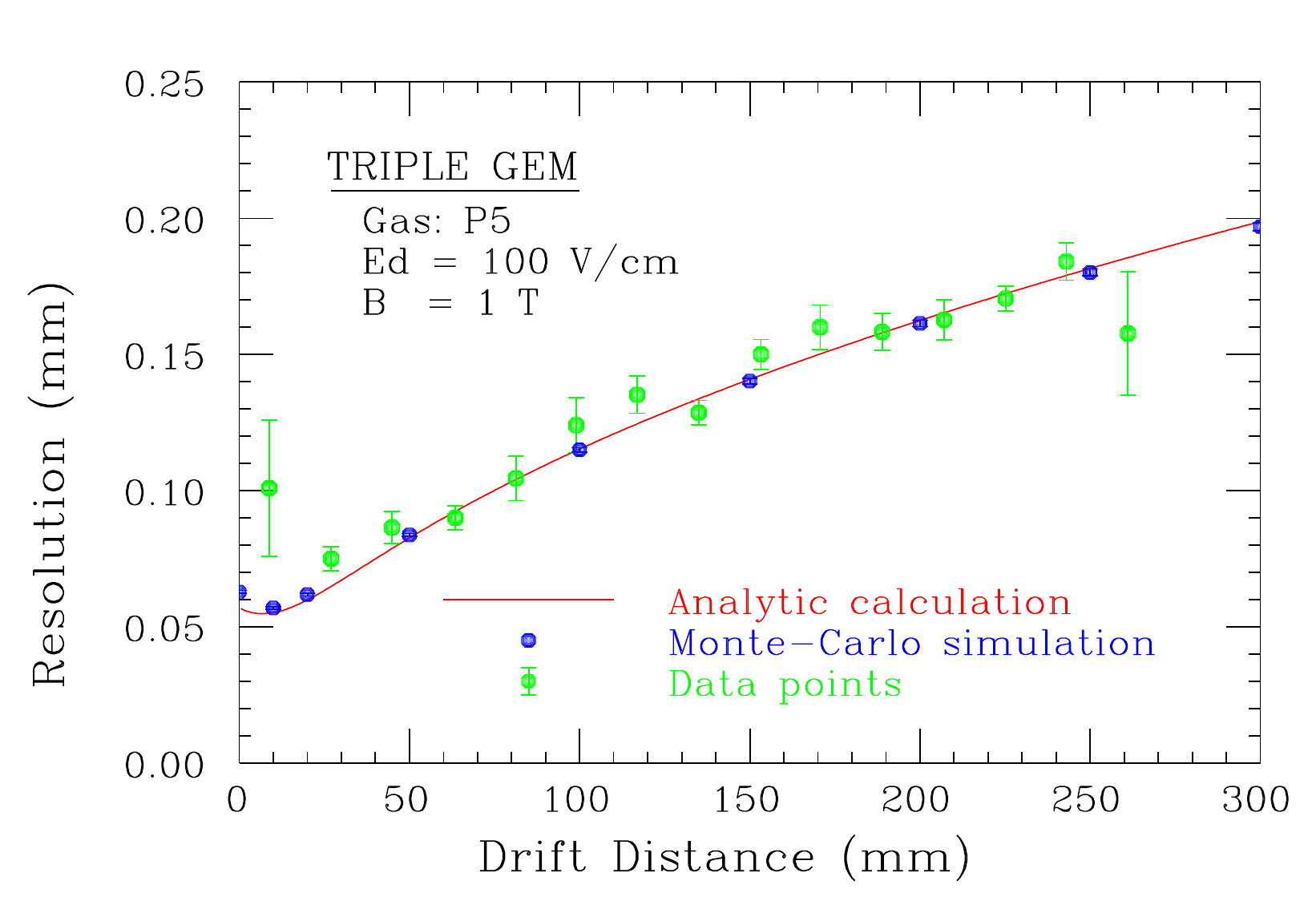}
\caption[figA1]{\label{figA3}
\footnotesize Comparison between the analytic calculation and the Monte-Carlo
simulation. In the calculation $N_{eff}$ is assumed to be 21
and PRF is assumed to be a Gaussian with $\sigma$ = 363 $\mu$m.
The diffusion constant ($D$) is set to 166 $\mu$m/$\sqrt{{\rm cm}}$ in both
cases. 
}
\end{figure}

Evaluation of Eq. (A.5) or (A.6), including the average over $\tilde{x}$,
can be done numerically using a short and simple program,
with much shorter demanded CPU time than Monte-Carlo simulations.
The results of the analytic calculation and a Monte-Carlo simulation are
compared in Fig.~\ref{figA3} for the triple GEM readout.
The Monte-Carlo simulation takes into account the primary ionization
statistics,
diffusion in the drift space, avalanche multiplication and its fluctuation
in the GEM holes,  
and the diffusion in the transfer and induction gaps.
The figure shows that they are almost identical, indicating the reliability of
both the analytic approach and the Monte-Carlo simulation.
A major advantage of Monte-Carlo simulation is that it can easily be
generalized to be applicable to inclined tracks.

To summarize,
the analytic calculation gives reliable evaluation of the spatial resolution
of a TPC for tracks perpendicular to the pad row once the effective number of
electrons ($N_{eff}$), the diffusion constant ($D$),
and the pad response function (PRF) are known. 
$N_{eff}$ is determined from the primary ionization statistics
(average density of primary ionizations and their cluster size distribution)
and the relative variance of avalanche fluctuation for
a single drift electron~\cite{Kobayashi}.
They are experimentally measurable or found in literature.
The diffusion constant in the drift region is determined from the slope of
the pad-response width squared as a function of drift distance (Eq. (A.2)).
It may be estimated using the simulation by MAGBOLTZ.
Finally, the width of pad response function is estimated from the intercept of
the squared pad-response width plotted against drift distance (Eq. (A.2)).
This can be estimated also by using the simulated value(s) of diffusion constant
in the detection gap(s).
The most reliable PRF would, however, be provided by a
dedicated experiment using a single-electron source and finer readout pads.  
%


\section*{\nonumber Acknowledgements}

The author would like to thank the people at Indian Institute of Science
for their support and hospitality.
He is also grateful to many colleagues in the ILC-TPC collaboration for their
continuous encouragement and support.

\end{document}